\documentclass[12pt]{article}
\usepackage{amsxtra,amssymb,amsthm,amsmath,latexsym}

\theoremstyle{plain}

\newtheorem{remark}{Remark}[section]

\def\oH{\buildrel\circ\over H}
\def\oH1{\buildrel\circ\over H\kern-.02in{}^1}

\begin{document}
\setcounter{page}{0}

In: Acoustic, Electromagnetic and Elastic wave scattering -- 
Focus on the $T$-matrix approach.

Pergamon Press, New York, 1980, pp. 537-546.
(Ed. V. Varadan and V. Varadan).


\title{ Wave scattering by small bodies of arbitrary shapes
\thanks{key words: wave scattering, small bodies of arbitrary
shapes }
\thanks{Math subject classification: 65J10, 65R20, 78A45    }
 }

\author{
  A.G. Ramm\\
  Mathematics Department, Kansas State University, \\
  Manhattan, KS 66506-2602, USA\\
  ramm@math.ksu.edu\\
  }

\date{}

\maketitle\thispagestyle{empty}


\section{Introduction}
The theory of wave scattering by small bodies was initiated by 
Rayleigh (1871). Thompson (1893) was the first to understand the
role of magnetic dipole radiation. Since then, many papers have
been published on the
subject because of its importance in applications. From a 
theoretical point
of view there are two directions of investigation:

\begin{description}
\item[(i)] to prove that the
scattering amplitude can be expanded in powers of $ka$, where
$k=2 \pi / \lambda$ and $a$ is a characteristic dimension of a small body,

\item[(ii)] to find the coefficients of the expansion efficiently.

\end{description}

Stevenson (1953),
Senior and Kleinman can be mentioned among 
contributors to the first topic.

To my knowledge, there were no results 
concerning the second topic for bodies
of an arbitrary shape. Such results are of interest in geophysics,
radiophysics, optics, colloidal chemistry and solid state theory.

In this paper we review the results of the author on 
the theory of scalar and vector wave scattering by small bodies
of an arbitrary shape with the emphasis on practical
applicability of the formulas obtained and on the mathematical
rigor of the theory.
For the scalar wave scattering by a single body, 
our main results can be described as follows:

\begin{description}
\item[(1)] Analytical formulas for the 
scattering amplitude for a small body of an
arbitrary shape are obtained; 
dependence of the scattering amplitude on the
boundary conditions is described.

\item[(2)]
An analytical formula for the scattering
matrix for electromagnetic wave scattering by a small 
body of an arbitrary
shape is given. Applications of these results are outlined 
(calculation of the properties of a rarefied medium; inverse
radio measurement problem;
formulas for the polarization tensors and capacitance).

\item[(3)] The multi-particle
scattering problem is analyzed and interaction 
of the scattered waves is taken
into account. For the self-consistent field in a medium consisting
of many particles $(\sim 10^{23})$, integral-differential 
equations are
found. The equations depend on the boundary conditions 
on the particle surfaces. These equations offer a 
possibility of solving the inverse problem
of finding the medium properties from the 
scattering data. For about 5 to 10
bodies the fundamental integral equations of the theory can be solved
numerically to study the interaction between the bodies.

\end{description}

In section 2 the results concerning the scalar 
wave scattering are described.
In section  3 the electromagnetic scattering 
is studied and the solution of the inverse
problem of radio measurements is outlined. 
In section 4 the many-body problem is examined.

\section{Scalar Wave Scattering by a Single Body}
Consider the problem
\begin{equation}
        \left( \nabla^2 + k^2 \right) v = 0 \hbox{\ in\ } \Omega; \quad
        \left(\frac{\partial v}{\partial N} - hv \right) \left|_{\Gamma}=
        -\left(\frac{\partial u_0}{\partial N} - hu_0 \right) \right|_{\Gamma},
        \end{equation}
\begin{equation}
        v \sim \frac{\exp \left( ik|x| \right)}{|x|} f(n,k)
        \hbox{\ as\ } |x| \to \infty, \quad
        \frac{x}{|x|} = n,
        \end{equation}
where $\Omega = R^3 \backslash D$, $D$ is a bounded domain with a smooth
boundary $\Gamma$, $N$ is the outer normal to $\Gamma$, $u_0$ is the initial
field which is usually taken in the form $u_0 = \exp$ $\{ik(\nu, x)\}$.
We look for a solution of the problem (1)-(2) of the form
\begin{equation}
        v = \int_\Gamma
        \frac{\exp \left(ikr_{xt}\right) \sigma (t)} {4 \pi r_{xt}} dt,
        \quad r_{xt} = |x-t|,
        \end{equation}
and for the scattering amplitude $f$ we have the formula
\begin{equation}
        f = \frac{1}{4 \pi} \int_\Gamma \exp \{-ik(n,t)\} \sigma (t,k)dt=
        \frac{1}{4 \pi} \int_\Gamma \sigma_0 (t)dt + O(ka),
        \end{equation}
where
\begin{equation}
        \sigma(t,k) = \sigma_0 (t) + ik\sigma_1 (t) +
        \frac{(ik)^2}{2} \sigma_2(t) + \dots
        \end{equation}
Putting (3) in the boundary conditions (1) we get the integral equation for
$\sigma$:
\begin{equation}
        \sigma = A(k) \sigma - hT(k) \sigma - 2hu_0 + 2
        \frac{\partial u_0}{\partial N},
        \end{equation}
where
\begin{equation}
        A(k) \sigma = \int_\Gamma \frac{\partial}{\partial N_s} \quad
        \frac{\exp \left(ikr_{st} \right)}{2\pi r_{st}} \sigma (t) dt, \quad
        T(k) \sigma = \int_\Gamma
        \frac{\exp \left(ikr_{st} \right)}{2 \pi_{st}} \sigma (t) dt.
        \end{equation}
Expanding $\sigma$, $A(k)$ and $T(k)$ in the powers $k$ and equating the
corresponding terms in (6) we obtain, for $h=0$, i.e. for the Neumann boundary
condition, the following equations:
\begin{equation}
        \sigma_0 = A_0 \sigma_0,
        \end{equation}
\begin{equation}
        \sigma_1 = A_0 \sigma_1 + A_1 \sigma_0 + 2 
        \frac{\partial u_{01}}{\partial N},
        \end{equation}
\begin{equation}
        \sigma_2 = A_0 \sigma_2 + 2A_1 \sigma_1 + A_2 \sigma_0 + 2
        \frac{\partial u_{02}}{\partial N},
        \end{equation}
etc, where
\begin{equation}
        A(k) = A_0 + ikA_1 + \frac{(ik)^2}{2} A_2 + \dots;
        \quad  u_0 =
        u_{00} + iku_{01} + \frac{(ik)^2}{2}u_{02}
        + \dots.
        \end{equation}
Expanding $f$ in formula (4) we obtain, up to the terms of the second order:
\begin{equation}
        \begin{array}{l}
        f \quad= \frac{1}{4 \pi} \int_\Gamma \sigma_0 dt + ik
        \left\{ \frac{1}{4 \pi} \int_\Gamma \sigma_1 dt +
        \left( n, \int_\Gamma \sigma_0 (t) tdt \right) \right\}
        \\ \\
        \quad + \frac{(ik)^2}{2}
        \left\{ \frac{1}{4 \pi} \int_\Gamma \sigma_2 dt +
        \frac{2}{4 \pi} \left( n, \int_\Gamma \sigma_1 tdt \right)
        + \frac{1}{4 \pi} \int_\Gamma \sigma_0 (n,t)^2dt \right\}.
        \end{array}
        \end{equation}
From (8) it follows that $\sigma_0 = 0$ and from (9) it follows that
$\int_\Gamma \sigma_1dt = 0$. Some calculations lead to the following final
result (see \cite{17}):
\begin{equation}
        f = \frac{ikV}{4 \pi} \beta_{pq}n_p \frac{\partial u_0}{\partial x_q}
        \left|_{x=0} + \frac{V}{4 \pi} \Delta u_0 \right|_{x=0}.
        \end{equation}
Usually $u_0 = \exp \{ik(\nu, x)\}$, and in this case formula (13) can be
written as:
\begin{equation}
        f = - \frac{k^2V}{4 \pi} \left( \beta_{pq} \nu_qn_p +1 \right),
        \tag{$13'$}\end{equation}
where over the repeated indices the summation is understood,
$V$ is the volume of the body $D$ and $\beta_{pq}$ is the
magnetic polarizability tensor of $D$. Note that $f \sim k^2 a^3$ the
scattering is anisotropic and is defined by the tensor $\beta_{pq}$.
Formula (25) below allows one to calculate $\beta_{pq}$.

For $h= \infty$ (the Dirichet boundary condition) integral equation (6) takes
the form:
\begin{equation}
        T(k)\sigma = -2u_0.
        \end{equation}
Hence
\begin{equation}
        \int_\Gamma \frac{\sigma_0 dt}{4\pi_{st}} = -u_0 \bigg|_\Gamma.
        \notag\end{equation}
Since $ka << 1$ the field $u_0 \left|_\Gamma = u_0(x,k) \right|_{x=0}$,
where the origin is assumed to be inside the body $D$. From the above
equation it follows,
that $\int_\Gamma \sigma_0 dt = -Cu_0$,
\begin{equation}
        f = -\frac{Cu_0}{4 \pi},
        \end{equation}
where $C$ is the capacitance of a conductor with the shape $D$. Hence for the
Dirichet boundary condition, $f \sim a$, where $a$ is a characteristic length
of $D$, and the scattering is isotropic.

For $h \neq 0$, using the same line of arguments, it is possible to obtain
the following approximate formula for the scattering amplitude:
\begin{equation}
        f \approx -\frac{hS}{4 \pi \left(1+hSC^{-1}\right)} \quad
        u_{00},
        \end{equation}
where $S= \hbox{\ meas\ }(\Gamma)$, i.e., the surface area of $\Gamma$,
and $C$ is the capacitance of $D$.
If $h$ is very
small $\left(h \sim k^2a^3 \right)$ the formula for $f$ should be changed
and the terms analogous to (13) should be taken into account.

\section{Electromagnetic Wave Scattering by a Single Body}
If a homogeneous body $D$ with the parameters $\varepsilon$, $\mu$, $\sigma$,
is placed into a homogeneous medium with the parameters $\varepsilon_0$,
$\mu_0$, $\sigma_0$, then the following formula for the scattering matrix
$\mathcal{S}$
was established by the author (see \cite{17}):
\begin{equation}
        \mathcal {S} = \frac{k^2V}{4 \pi}
        \left[ \begin{array}{l}
        \mu_0 \beta_{11} + \alpha_{22} \cos \theta - \alpha_{32} \sin \theta,
        \quad \alpha_{21} \cos \theta - \alpha_{31} \sin \theta - \mu_0
        \beta_{12} \\
        \alpha_{12}- \mu_0 \beta_{21} \cos \theta + \mu_0 \beta_{31}
        \sin \theta, \quad \alpha_{11} + \mu_0\beta_{22} \cos \theta -
        \mu_0 \beta_{32} \sin\theta
        \end{array}\right]_,
        \end{equation}
where $\mathcal {S}$ is defined by the formula
\begin{equation}
        \left( \begin{array}{l}
        f_2 \\ f_1
        \end{array} \right)
        = \mathcal {S}
        \left( \begin{array}{l}
        E_2 \\ E_1
        \end{array} \right)
        =
        \left( \begin{array}{l}
        S_2 S_3 \\ S_4 S_1
        \end{array} \right)
    \left( \begin{array}{l}
        E_2 \\ E_1
        \end{array} \right)_,
        \end{equation}
$\theta$ is the angle of scattering, $E_1$, $E_2$ are the components of the
initial field, $f_1$, $f_2$ are the components of the scattered field in the
far field region multiplied by 
$|x|^{-1} \exp (ik|x|)$, the plane YOZ is the plane
of scattering,
$\alpha_{ij} = \alpha_{ij}(\gamma)$,
is the polarizability tensor, $\gamma = \left(\varepsilon - \varepsilon_0
\right) / \left(\varepsilon + \varepsilon_0 \right)$
and
$\beta_{ij} = \alpha_{ij} (-1)$ is the magnetic polarizability tensor.

If one knows $\mathcal S$ one can find all the values of interest
to physicists for
electromagnetic wave propagation in a rarefied medium consisting of small
bodies. The tensor of refraction coefficient can be calculated by the formula
$n_{ij} = \delta_{ij} + 2 \pi Nk^{-2} S_{ij} (0)$, where $N$ is the number
of bodies per unit volume. The tensor
$\alpha_{ij}(\gamma)$ can be calculated analytically by the formula
\begin{equation}
        \left| \alpha_{ij} (\gamma) - \alpha_{ij}^{(n)} (\gamma) \right| \leq
        Aq^n, \quad 0 < q < 1,
        \end{equation}
where $A$, and $q$ are some constants depending only on the geometry of the
surface, and
\begin{equation}
        \alpha^{(n)}_{ij}: = \frac{2}{V} \sum^n_{m=0}
        \frac{(-1)^m}{(2 \pi)^m} \quad
        \frac{\gamma^{n+2} - \gamma^{m+1}}{\gamma -1} 
        b^{(m)}_{ij}, \quad
        n \geq 1.
        \end{equation}
In (20)
\begin{equation}
        b^{(0)}_{ij} = V\delta_{ij},
        \quad b^{(1)}_{ij} =
        \int_\Gamma \int_\Gamma
        \frac{N_i(s)N_j(t)dsdt}{r_{st}},
        \end{equation}
\begin{equation}
     \begin{array}{c}
        b^{(m)}_{ij} = \int_\Gamma \int_\Gamma dsdt N_i (s) N_j (t)
        \underbrace{\int_\Gamma\dots \int_\Gamma}_{m-1}
        \frac{1}{r_{st}} \psi \left(t_1, t \right) \dots
        \psi \left(t_{m-1}, t_{m-2} \right) dty \dots
       \\ \\
        dt_1 \dots dt_{m-1}; \quad \psi(t,s) \equiv
        \frac{\partial}{\partial N_t} \quad
        \frac{1}{r_{st}}.
       \end{array}
       \end{equation}
In particular
\begin{equation}
        \alpha^{(1)}_{ij} (\gamma) = 2 \left(\gamma + \gamma^2 \right)
        \delta_{ij} - \frac{\gamma^2 b^{(1)}_{ij}}{\pi V}, \quad
        \beta^{(1)}_{ij} = -\frac{b^{(1)}_{ij}}{\pi V}.
        \end{equation}
For particles with $\mu = \mu_0$ and $\varepsilon$ not very large, so that the
depth $\delta$ of the skin layer is considerably larger than $a$, one can
neglect the magnetic dipole radiation and in formula (17) for the scattering
matrix one can omit the terms with the multipliers $\beta_{ij}$.

The vectors of electric $P$ and magnetic $M$ polarizations can be found by the
following formulas, respectively,
\begin{equation}
        P_i = \alpha_{ij} (\gamma)V \varepsilon_0 E_j, \quad \gamma :=
        \frac{\varepsilon - \varepsilon_0}{\varepsilon + \varepsilon_0},
        \end{equation}
where $E_j$ is the initial field, over the repeated indices one sums up, and
\begin{equation}
        M_i = \alpha_{ij} (\widetilde{\gamma}) V \mu_0 H_j, \quad
        + \beta_{ij} V \mu_0 H_j, \quad
        \widetilde{\gamma} :=
        \frac{\mu - \mu_0}{\mu + \mu_0}, \quad
        \beta_{ij} := \alpha_{ij} (-1),
        \end{equation}
where $H_j$ is the initial field and the second term on the right hand side of
equality (25) should be omitted if the skin-layer depth
$\delta >> a$.

The scattering amplitudes can be found from the formulae
\begin{equation}
        f_E = \frac{k^2}{4 \pi \varepsilon_0} \left[n,[P,n]\right] +
        \frac{k^2}{4 \pi} \sqrt{\frac{\mu_0}{\varepsilon_0}}[M,n],
        \end{equation}
\begin{equation}
        f_H = \sqrt{\frac{\varepsilon_0}{\mu_0}}[n, f_E],
        \end{equation}
where $[A,B]$ stands for the vector product $A \times B$, and $P$ and $M$
can be
calculated by formulae (24), (25), (19), (20), (22). If $\delta >> a$
one can neglect the second term on the right-hand side of (26).

It is possible
to give a simple solution to the following inverse problem which can be called
the inverse problem of radiomeasurements.

Suppose an initial electromagnetic
field is scattered by a small probe. Assume that the scattered field
$E^\prime$, $H^\prime$ can be measured in the far field region.
The problem is:
calculate the initial field at the point where the small probe detects
$E^\prime$, $H^\prime$. This problem is of interest, for example, when one
wants to determine the
electromagnetic field distribution in an antenna's aperture. Let us assume for
simplicity that for the probe $\delta >> a$, so that
\begin{equation}
        E^\prime = \frac{\exp(ikr)}{r} \frac{k^2}{4 \pi \varepsilon_0}
        \left[n[P,n]\right].
        \end{equation}

From (28) one can find
$P-n_1 \left(P,n_1 \right) = E^\prime \left(n_1 \right)b$
where
$b = \frac{\exp(ikr)}{r}$ $\frac{k^2}{4 \pi \varepsilon_0}$.
A measurement in the $n_2$ direction,
where $\left(n_1, n_2 \right) = 0$,
results in $P-n_2(P,n_2)= E^\prime(n_2)b$.
Hence $(n_1,P) = b(E^\prime(n_2), n_1)$.
Thus $P = b\{E^\prime (n_1) + n_1(E^\prime(n_2), n_1)\}$.
But
\begin{equation}
        P_i = \alpha_{ij} (\gamma) V \varepsilon_0 E_j.
        \tag{$\ast$}
        \end{equation}
Since $V$ and $\varepsilon_0$ are known and $\alpha_{ij}(\gamma)$ can be
calculated by formulae (19), (20) and the matrix $\alpha_{ij}$ is positive
definite (because $\frac{1}{2} \alpha_{ij} V \varepsilon_0 E_jE_i$ is the
energy) it follows that system $(\ast)$ is uniquely solvable. Its solution
is the desired vector $E$.

Let us give a formula for the capacitance of a conductor $D$ of an arbitrary
shape, which proved to be very useful in practice:

\begin{equation}
        C^{(n)} = 4 \pi \varepsilon S^2
        \left\{ \frac{(-1)^n}{(2 \pi)^n} \int_\Gamma \int_\Gamma
        \frac{dsdt}{r_{st}}
     \underbrace{\int_\Gamma\dots
     \int_\Gamma}_{n \hbox{\begin{tiny}\ times\end{tiny}}}
        \psi(t,t_1) \dots \psi (t_{n-1}, t_n) dt_1 \dots dt_n
        \right\}^{-1}_,
        \end{equation}
\begin{equation}
        C^{(0)} = \frac{4 \pi \varepsilon_0 S^2}{J} \leq C, \quad
        J \equiv \int_\Gamma \int_\Gamma \frac{dsdt}{r_{st}}, \quad
        S = \hbox{meas} \Gamma.
        \end{equation}

It can be proved that
\begin{equation}
        \left|C - C^{(n)}\right| \leq Aq^n, \quad 0 < q < 1,
        \tag{$30'$}\end{equation}
where $A$ and $q$ are constants which depend only on the geometry of $\Gamma$.

\begin {remark}
The theory is also applicable for small layered bodies.
\end{remark}
\begin {remark}
Two sided variational estimates for $\alpha_{ij}$ and $C$ were given in
\cite{17} and \cite{18}.
\end{remark}
\section {Many Body Wave Scattering}
First we describe a method for solving the scattering problem for $r$
bodies, $r \sim 5-10$, and then we derive an integral-differential equation
for the self-consistent field in a medium consisting of many
$(r \sim 10^{23})$ small bodies. We look for a solution of the scalar wave
scattering problem of the form
\begin{equation}
        u = u_0 + \sum^r_{j=1} \int_{\Gamma_j}
        \frac{\exp(ikr_{st})}{4 \pi r_{xt}} \sigma_j (t)dt.
        \end{equation}
Applying the boundary condition,
\begin{equation}
        u \bigg|_{\Gamma_j} = 0, \quad 1 \leq j \leq r,
        \end{equation}
we obtain the system of $r$ integral equations for the $r$ unknown functions
$\sigma_j$. In general this system can be solved numerically. When
$d << \lambda$, where $d= \hbox{min}_{i \neq j} d_{ij}$, and $d_{ij}$
is the distance between $i-th$ and $j-th$ body, the system of the integral
equations has dominant diagonal terms and it can be easily solved by an
iterative process, the zero approximation being the initial field
$u_0$.

If $ka >>1$, $d >>a$, but not necessarily $d >> \lambda$, the average
field in the medium consisting of small particles can be found from the
integral
equation
\begin{equation}
        u(x,k) = u_0(x) - \int_{R^3} \frac{\exp(ikr_{xy})}{4 \pi r_{xy}}
        q(y) u (y,k)dy.
        \end{equation}

Here $q(y)$ is the average value of
$h_jS_j \left(1+h_jS_jC_j^{-1} \right)^{-1}$
over the volume $dy$ in a neighborhood of $y$ for bodies with impedance
boundary conditions. For $h_j = \infty$ (the Dirichlet boundary condition)
and identical bodies, one has
$q(y)=N(y)C$, where $N(y)$ is the number of the bodies per unit volume and
$C$ is the capacitance of a body. For the Neumann boundary condition the
corresponding equation is the integral-differential equation:
\begin{equation}
        u(x,k) = u_0(x,k) + \int_{R^3} \frac{\exp(ikr_{xy})}{4 \pi r_{xy}}
        \left\{ B_{pq} (y) \frac{\partial u(y,k)}{\partial y_q} \quad
        \frac{x_p - y_p}{r_{xy}} + \frac{1}{2} \Delta u(y,k) b(y) \right\} dy,
        \end{equation}
where
\begin{equation}
        b(y) = N(y)V, \quad B_{pq}(y) = ikV \beta_{pq} N(y),
        \end{equation}
$V$ is the volume of a body, and $\beta_{pq}$ is its magnetic polarizability
tensor (see forumla (25)). The solution to equations (33) and (34)
can be considered as the
self-consistent (effective) field acting in the medium.

Equations (33) and (34) allow one to solve the inverse problems of the
determination of the medium properties from the scattering data. For example,
from (33) it follows that the scattering amplitude has the form
\begin{equation}
        f = -\frac{1}{4 \pi} \int \exp \{-ik(n,y)\}q(y) u(y,k)dy.
        \end{equation}
For a rarefied medium it is reasonable to replace $u$ by $u_0$
(the Born approximation) and to obtain
\begin{equation}
        f \approx -\frac{1}{4 \pi} \int_{R^3} \exp \{-ik(n,y)\}q(y) u_0
        (y,k)dy.
        \end{equation}
If $u_0 = \exp\{ik(\nu, x) \}$ formula (37) is valid for $k>>1$ with the
error $O(k^{-1})$ if
\begin{equation}
        \left|q(x)\right| + \left| \nabla q(x) \right| \leq
        \frac{c}{1 + |x|^{3 + \varepsilon}}, \quad
        \varepsilon >0.
        \notag\end{equation}

Hence if $f$ is known for all $0<k< \infty$, and all $(n- \nu) \in S_2$,
where $S_2$ is the unit
sphere in $R^3$, the Fourier transform of $q(y)$ is known, and $q$ can be
uniquely determined. If $q(y)$ is compactly supported,
 i.e. $q(y)=0$ outside
some bounded domain, then $f$ is an entire function of $k$,
and knowing $f$ in any interval
$[k_0, k_1]$, $0<k_0 < k_1$, for all $(n- \nu) \in S_2$ one can find
$f$ for all $0<k< \infty$ uniquely by analytic continuation, and thus one can
determine $q(y)$ uniquely.

Let us consider the $r$-body problem for a few bodies, $r \sim 10$
(small $r$).
Assume that the
Dirichlet boundary condition holds. Let us look for a solution of the form
\begin{equation}
        u(x) = u_0 + \sum^r_{j=1} \int_{\Gamma_j}
        \frac{\exp(ik|x-t|)}{4 \pi |x-t|} \sigma_j (t)dt.
        \end{equation}
The scattering amplitude is equal to
\begin{equation}
        f(n,k) = \frac{1}{4 \pi} \sum^r_{j=1} \exp \left\{ -ik(n,t_j) \right\}
        \int_{\Gamma_j} \exp \left\{ -k(n,t-t_j) \right\}
        \sigma_j(t)dt,
        \end{equation}
where $t_j$ is some point inside the j-th body. Since $ka<<1$ this formula
can be rewritten as:
\begin{equation}
        f(n,k) = \frac{1}{4 \pi} \sum ^r_{j=1} \exp \left\{-ik(n,t_j) \right\}
        Q_j,
        \end{equation}
where
\begin{equation}
        Q_j = \int_{\Gamma_j} \sigma_{j0} dt + O(ka), \quad
        \sigma_{j0} = \sigma_j \bigg|_{k=0}.
        \end{equation}
This is the same line of arguments as in Section 2.
Using the boundary condition one gets:
\begin{equation}
        \sum^r_{j \neq m, j=1} \int_{\Gamma_j}
        \frac{\exp(ik|x_m-t|)}{4 \pi |x_m-t|} \sigma_j(t) dt +
        \int_{\Gamma_m} \frac{\exp(ik|x_m-t|)}{4 \pi|x_m-t|}
        \sigma_mdt = -u_0(x_m).
        \end{equation}
With the accuracy of $O(ka)$, this can be written as:
\begin{equation}
        \int_{\Gamma_m} \frac{\sigma_m(t)dt}{4 \pi|x_m-t|} +
        \sum^r_{j=1, j \neq m} \frac{\exp(ikd_{mj})}{4 \pi d_{mj}}
        Q_j = -u_{0m}, \quad
        1 \leq m \leq r,
        \end{equation}
where $d_{mj} = |x_m-t_j|$. If $C_m$ is the capacitance of the m-th body one
can rewrite (43) as:
\begin{equation}
        Q_m = -C_mu_{0m} - \sum^r_{j=1, j \neq m} C_m
        \frac{\exp(ikd_{mj})}{4 \pi d_{mj}} Q_j, \quad
        1 \leq m \leq r.
        \end{equation}
This is a linear system from which $Q_j$, $1 \leq m \leq r$, can be found.
If $d_{mj}C^{-1}_m >>1$, this system can be easily solved by an iterative
process. If $\{Q_j\}$ are known, then the scattering amplitude can be found
from (40).

More details about the described theory the reader can find in the
References, especially in he monograph \cite{17}.


\begin{thebibliography}{100}

\bibitem{1} 
A.G. Ramm,
``Iterative solution of integral equation in potential theory",
Sov. Phys. Dokl., 186, (1969), 62-65. {\it Math Rev.}, 41 \#9462.

\bibitem{2}
A.G. Ramm,
``Approximate formulas for tensor polarizability and capacitance of bodies
of arbitrary shape and its applications", Sov. Phys. Dokl., 195,
(1970), 1303-1306. {\it Math. Rev.} 55 \#1947

\bibitem{3}
A.G. Ramm,
``Calculation of the initial field from scattering amplitude",
{\it Radiotech. i Electron.}, 16, (1971), 554-556.

\bibitem{4}
A.G. Ramm,
``Approximate formulas for polarizability tensor and capacitances for bodies
of an arbitrary shape", Radiofisika, 14, (1971), 613-620. {\it Math Rev.}
47 \#1386.

\bibitem{5}
A.G. Ramm,
``Electromagnetic wave scattering by small bodies of an arbitrary shape",
Proc. Fifth all Union Sumpos. on Wave Diffraction, {\it Trudy Math. Inst.}
Steklova, Leningrad, (1971), 176-186.

\bibitem{6}
A.G. Ramm,
``Calculation of the magnetization of thin films", Microelctronical 6, (1971),
65-68. (with Frolov).

\bibitem{7}
A.G. Ramm,
``Calculation of scattering amplitude of electromagnetic waves by small
bodies of an arbitrary shape II.", Radiofisika, 14, (1971), 1458-1460.

\bibitem{8}
A.G. Ramm,
``Electromagnetic wave scattering by small bodies of an arbitrary shape
and relative topics", Proc. Intern. Sympos. URSI, Moscow, (1971), 536-540.

\bibitem{9}
A.G. Ramm,
``Calculation of the capacitance of a parallelpiped", Electricity, 5, (1972),
90-91. (with Golubkova, Usoskin).

\bibitem{10}
A.G. Ramm,
``On the skin-effect theory", {\it Journ. of Techn. Phys.}, 42, (1972),
1316-1317.

\bibitem{11}
A.G. Ramm,
``Calculation of the capacitance of a conductor placed in anisotropic
inhomogeneous dialectric", Radiofisika, 15, (1972), 1268-1270.
{\it Math. Rev.} 47 \# 2284.

\bibitem{12}
A.G. Ramm,
``Remark to integral equation theory", Diff eq., 8 (1972), 1517-1520.
Engl. transl. pp.1117-1180; {\it Math. Rev.} 47 \#2284.

\bibitem{13}
A.G. Ramm,
``Iterative process to solve the third boundary problem", Diff. eq., 9, (1973),
2075-2079. {\it Math. Rev.} 48 \#6861.

\bibitem{14}
A.G. Ramm,
``Light scattering matrix for small particles of an arbitrary shape",
Opt. and Spectroscopy, 37, (1974), 125-129.

\bibitem{15}
A.G. Ramm,
``Scalar scattering by the set of small bodies of an arbitrary shape",
Radiofisika, 17, (1974), 1062-1068.

\bibitem{16}
A.G. Ramm,
``New Methods of calculation of the static and quasistatic electromagnetic
waves", Proc. Fifth Int. Symp. ``Radio-electronics-74", Sofia, 3,
(1974), 1-8 (report 12).

\bibitem{17}
A.G. Ramm,
``Iterative methods for calculating the static fields and wave scattering
by small bodies", Springer Verlag, New York, 1982.

\bibitem{18}
A.G. Ramm,
``Estimates of some functionals in
quasistatic  electrodynamics", Ukrain. Phys. J. 5, 1975, 534-543.
 {\it Math.rev.}  56 \#14165.


\bibitem{19}
A.G. Ramm,
``Calculation of the scattering amplitude for
the wave scattering from small bodies of an arbitrary
  shape", Radiofisika, 12, 1969, 1185-1197.  {\it Math.rev.} 43
\#7131.



\end{thebibliography}
\end{document}